\documentclass[showpacs,prl,onecolumn,aps,superscriptaddress,preprintnumbers,nofootinbib]{revtex4}
\usepackage{amsmath,amssymb}
\usepackage{epsfig}
\usepackage{graphicx}
\usepackage{mathrsfs}
\usepackage{amsmath}
\usepackage{amsfonts}
\usepackage{epstopdf}
\usepackage{color}
\def\slashchar#1{\setbox0=\hbox{$#1$}     		% set a box for #1
   \dimen0=\wd0                                 	% and get its size
   \setbox1=\hbox{/} \dimen1=\wd1               	% get size of /
   \ifdim\dimen0>\dimen1                        	% #1 is bigger
      \rlap{\hbox to \dimen0{\hfil/\hfil}}      	% so center / in box
      #1                                        	% and print #1
   \else                                        	% / is bigger
      \rlap{\hbox to \dimen1{\hfil$#1$\hfil}}   	% so center #1
      /                                         	% and print /
   \fi}

\renewcommand{\vec}{\boldsymbol}
\newcommand{\be}{\begin{equation}}
\newcommand{\ee}{\end{equation}}
\newcommand{\bea}{\begin{eqnarray}}
\newcommand{\eea}{\end{eqnarray}}
\newcommand{\ba}{\begin{array}}
\newcommand{\ea}{\end{array}}

\def\eq#1{{Eq.~(\ref{#1})}}
\def\fig#1{{Fig.~\ref{#1}}}
\newcommand{\bas}{\bar{\alpha}_S}
\newcommand{\as}{\alpha_S}
\newcommand{\nn}{\nonumber}

\newcommand{\Lb}{\left(}
\newcommand{\Rb}{\right)}

\begin{document}

%\title{Parton densities without partons}
\title{Deep inelastic scattering as a probe of entanglement}

\author{Dmitri E. Kharzeev}
\email{Dmitri.Kharzeev@stonybrook.edu}
\affiliation{Department of Physics and Astronomy, Stony Brook University, New York 11794-3800, USA}
\affiliation{Department of Physics and RIKEN-BNL Research Center, \\
Brookhaven National Laboratory, Upton, New York 11973-5000, USA}
\author{Eugene M. Levin}
\email{leving@post.tau.ac.il, eugeny.levin@usm.cl}
\affiliation{Department of Particle Physics, School of Physics and Astronomy,
Tel Aviv University, Tel Aviv, 69978, Israel}
\affiliation{Departamento de F\'\i sica,
Universidad T$\acute{e}$cnica Federico Santa Mar\'\i a   and
Centro Cient\'\i fico-Tecnol$\acute{o}$gico de Valpara\'\i so,
Casilla 110-V,  Valparaiso, Chile}

\date{\today}


\begin{abstract}
Using non-linear evolution equations of QCD, we compute the von Neumann entropy of the system of partons resolved by deep inelastic scattering at a given Bjorken $x$ and momentum transfer $q^2 = - Q^2$.  We interpret the result as the entropy of entanglement between the spatial region probed by deep inelastic scattering and the rest of the proton. At small $x$ the relation between the entanglement entropy $S(x)$ and the parton distribution $xG(x)$ becomes very simple: 
$S(x) = \ln[ xG(x) ]$. 
In this small $x$, large rapidity $Y$ regime, all partonic micro-states have equal probabilities -- the proton is composed by an exponentially large number $\exp(\Delta Y)$ of micro-states that occur with equal and exponentially small probabilities $\exp(-\Delta Y)$, where $\Delta$ is defined by $xG(x) \sim 1/x^\Delta$. For this equipartitioned state, the entanglement entropy is maximal -- so at small $x$, deep inelastic scattering probes a {\it maximally entangled state}. We propose the entanglement entropy as an observable that can be studied in deep inelastic scattering. This will require event-by-event measurements of hadronic final states, and would allow to study the transformation of entanglement entropy into the Boltzmann one. We estimate that the proton is represented by the maximally entangled state at $x \leq 10^{-3}$; this kinematic region will be amenable to studies at the Electron Ion Collider.
\end{abstract}
\maketitle

%%%%%%%%%%%%%%%%%%%%%%%%%%%%%%%%%%%%%%%%%%%%%%%%%%%%
\section{Introduction and summary}

In almost fifty years that ensued after the birth of the parton model \cite{Bjorken:1968dy,Feynman:1969ej,Feynman:1969wa,FEYN,BJ,Gribov},
   it has become an indispensable building block of high energy physics. The picture of ``quasi-free" partons ``frozen" in the infinite-momentum frame due to the Lorentz dilation is clear and intuitively appealing. Even more importantly, the parton model combined with the QCD factorization \cite{Collins:1989gx} allows to describe a vast variety of hard processes in terms of universal parton distributions. The renormalization group (RG) flow of QCD \cite{Gross:1973id,Politzer:1973fx,DeRujula:1974mnv} in terms of the parton model can be recast in the form of parton splitting \cite{Gribov:1972ri,Altarelli:1977zs,Dokshitzer:1977sg}, leading to the evolution of parton densities in Bjorken $x$ and virtuality $Q^2$. 
\vskip0.3cm
Nevertheless, in spite of a spectacular success of the parton model, it raises a number of conceptual questions:
\begin{itemize}
\item{The hadron in its rest frame is described by a pure quantum mechanical state $|\psi\rangle$ with density matrix $\hat{\rho} = |\psi\rangle \langle \psi |$ and zero von Neumann entropy $S = - \rm{tr} \left[ \hat{\rho} \ln \hat{\rho} \right] = 0$. How does this pure state evolve to the set of ``quasi-free" partons in the infinite-momentum frame? If the partons were truly free and thus incoherent, they would be characterized by a non-zero entropy. Since the Lorentz boost cannot transform a pure state into a mixed one, what is the precise meaning of ``quasi-free"? What is the rigorous definition of the parton distribution when applied to a pure quantum state? }

\item{Deep inelastic scattering (DIS) at Bjorken $x$ and momentum transfer $q^2 = - Q^2$ probes only a part of the proton's wave function; let us denote it $A$. In the proton's rest frame, where it is definitely described by a pure quantum mechanical state, the DIS probes the spatial region $A$ localized within a tube of radius $\sim 1/Q$ and length $\sim 1/(mx)$ \cite{Gribov:1965hf,Ioffe:1969kf}, where $m$ is the proton's mass. The inclusive DIS measurement thus sums over the unobserved part of the wave function localized in the region $B$ complementary to $A$, so we have access only to the reduced density matrix $\hat{\rho}_A = \rm{tr}_B \hat{\rho}$, and not the entire density matrix $\hat{\rho} = |\psi\rangle \langle \psi |$. Is there an entanglement entropy $S_A = - \rm{tr} \left[ \hat{\rho}_A \ln \hat{\rho}_A \right]$ associated with the DIS measurement? If there is, how does it relate to the conventional parton distribution? 
}

\item{What is the relation between the parton distribution and the multiplicity of final state hadrons in deep inelastic scattering? Is the ``parton liberation" \cite{Mueller:1999fp} picture universal, or does it apply only in the parton saturation domain? What is the interpretation of parton saturation \cite{Gribov:1984tu} and color glass condensate \cite{McLerran:1993ni,McLerran:1993ka,Gelis:2010nm} in terms of von Neumann entropy?
}

\end{itemize}

Answering these questions would also allow to interpret the inelastic electron scattering measurements in the domain of strong coupling (relevant at small and moderate $Q^2$), where the concept of quasi-free partons does not apply. Even at very large momentum transfer, this domain is relevant in the interpretation of deep inelastic scattering --  this is because the parton evolution equations describing the renormalization group (RG) flow in QCD depend on the initial conditions at some moderate initial $Q^2 = Q_0^2$. There is another reason for relating parton distributions to the entanglement entropy -- there exist quantum bounds on entropy (see for example \cite{Bekenstein:1980jp,Bousso:1999xy,Ryu:2006bv,Hubeny:2007xt}), whereas {\it a priori} there is no bound on the growth of parton distributions\footnote{The growth of the total cross section is limited by the Froissart theorem; however the relation of parton distributions to the cross section gets modified in the domain of high parton densities due to shadowing corrections to the scattering amplitude. Let us emphasize from the beginning that here we will view the parton distribution as the multiplicity of partons at a given $x$ and $Q^2$.} at small $x$ and large $Q^2$.
\vskip0.3cm

In this paper we attempt to address these questions in the framework of high energy QCD, see \cite{KOLEB} for an introduction. Before we proceed to presenting the derivation, let us state our main results:

\begin{enumerate}
\item{Using both a toy $(1+1)$ dimensional model of non-linear QCD evolution and full non-linear $(3+1)$ dimensional evolution equations,  we have computed the von Neumann entropy of partons $S(x)$ at a given $x$ (and $Q^2$, for the $(3+1)$ case) -- it is given by equations (\ref{TM8}) and (\ref{PCS5}).
}
\item{We have found the relation between the von Neumann entropy $S(x)$ and the gluon distribution\footnote{Here the parton distribution $xG(x)$ is defined as the number of gluons at a given $x$.} $xG(x)$ accessed in deep inelastic scattering. At small $x$ this relation becomes very simple:
\be\label{ent_part}
S(x) = \ln[ xG(x) ] .
\ee
Equation (\ref{ent_part}) implies that all microstates of the system are equally probable, and the von Neumann entropy is maximal. We argue that this equipartitioning of microscopic states that maximizes the von Neumann entropy corresponds to the parton saturation.
}
\item{At small $x$, we find that the von Neumann entropy diverges logarithmically at small $x$:
\be\label{sb1}
S(x) = \Delta \ln[1/x] = \Delta \ln \frac{L}{\epsilon},
\ee
where $L = (mx)^{-1}$ is the longitudinal distance probed in DIS ($m$ is the proton mass) and $\epsilon \equiv 1/m$ is the proton's Compton wavelength, see Fig.\ref{casc}; $\Delta$ is defined by $xG(x) \sim 1/x^\Delta$. This expression reminds the well known result for the entanglement entropy in  $(1+1)$ conformal field theory (CFT) \cite{HLW,CARDY}
\be\label{se1}
S_E = \frac{c}{3} \ \ln\frac{L}{\epsilon},
\ee
where $L$ is the length of the studied region, $\epsilon$ is the regularization scale describing the resolution of the measurement, and $c$ is the central charge of CFT that counts the number of degrees of freedom. We argue that this agreement is not coincidental, and propose  that the parton distributions, and the  entropy associated with them, arise from the entanglement between the spatial domain probed by DIS and the rest of the target.  Therefore the maximal value of the entanglement entropy attained at small $x$ implies that the corresponding partonic state is {\it maximally entangled}. Unlike the parton distribution, the entanglement entropy is an appropriate observable even at strong coupling when the description in terms of quasi-free partons fails.
}
\item{Assuming that the second law of thermodynamics applies to entanglement entropy (see e.g. \cite{Hubeny:2007xt} for a discussion), we get an inequality for the entropy of final state hadrons $S_h$ and the entropy of the initial state $S(x)$ in DIS:  $S_h \geq S(x)$. There are indications from holography that the entropy may not increase in the real-time evolution of strongly coupled systems (for a recent result, see \cite{Megias:2016vae}); this would imply the proportionality $S_h \sim S(x)$ in accord with the ``parton liberation" picture \cite{Mueller:1999fp} and ``local parton-hadron duality" \cite{Dokshitzer:1987nm}. This relation can be tested in deep inelastic scattering experiments by using event-by-event measurements of hadronic final state.
}
\end{enumerate}

The entanglement entropy between the large $x$ and small $x$ components of the wave function due to QCD evolution has recently been addressed in Ref. \cite{Kovner:2015hga}, see also \cite{Peschanski:2012cw}. This ``momentum space entanglement" \cite{Balasubramanian:2011wt} characterizing the renormalization group flow is different from the entanglement between the spatial region probed by deep inelastic scattering and the rest of the proton that is the object of our present study.
%%%%%%%%%%%%%%%%%%%%%%%%%%%%%%%%%%%%%%%%%%%%%%%%%%%%
\section{Entanglement entropy and parton distributions}
%%%%%%%%%%%%%%%%%%%%%%%%%%%%%%%%%%%%%%%%%%%%%%%%%%%%%%%%%%%%%%%%%%%%%%%%%%%%%%%%%%%%%%%%%%%%%%%%%%
%%%%%%%%%%%%%%%%%%%%%%%%%%%%%%%%%%%%%%%%%%%%%%%%%%%%

\subsection{Quantum mechanics of parton entanglement}

The entropy $S$ of a macroscopic state is given by 
the logarithm of the number $W$ of distinct microscopic states that compose it -- the Boltzmann formula $S = k \log W$ forms the basis of statistical physics (and is appropriately inscribed on Boltzmann's tombstone). 
Since partons are introduced as the microscopic constituents  that compose the macroscopic state of the proton, it seems natural to evaluate the corresponding entropy. 
However, the proton as a whole is a pure quantum state with a zero von Neumann entropy -- so we come to the apparent contradiction described in the Introduction.
\vskip0.3cm

To resolve it, let us consider a  region of space $A$ (for simplicity of notation, one-dimensional) probed by deep inelastic scattering (DIS). In the proton's rest frame, it is a segment of length $L = (mx)^{-1}$, where $m$ is the proton mass and $x$ is the Bjorken $x$ \cite{Gribov:1965hf,Ioffe:1969kf}. Let us denote by $B$ the region of space complementary to $A$, so that the entire space is $A \cap B$. The physical states inside the region $A$ probed by DIS are states in a Hilbert space $\mathcal{H}_A$ of dimension $n_A$, and unobserved states in the region $B$ belong to the Hilbert space $\mathcal{H}_B$ of dimension $n_B$. The composite system in $A \cap B$ (the entire proton) is then described by the vector $|\Psi_{AB}\rangle$ in the space $\mathcal{H}_A \otimes \mathcal{H}_B$ that is a tensor product of the two spaces:
\be\label{vecspace}
|\Psi_{AB}\rangle = \sum_{i,j} c_{ij}\ |\varphi_i^A\rangle \otimes |\varphi_j^B\rangle ,
\ee
where $c_{i,j}$ are the elements of the matrix $C$ that has a dimension $n_A \times n_B$. If one can find such states $|\varphi^A\rangle$ and $|\varphi^B\rangle$ that $|\Psi_{AB}\rangle = |\varphi^A\rangle \otimes |\varphi^B\rangle$, i.e. that the sum (\ref{vecspace}) contains only one term, then the state $|\Psi_{AB}\rangle$ is separable, or a product state. Otherwise the state $|\Psi_{AB}\rangle$ is entangled.
\vskip0.3cm
Let us introduce the coordinates $y \in A$ and $z \in B$. 
%\footnote{For simplicity we denote by coordinate $y$ the coordinate of all constituents $\in A$, as well as $z$   stands for coordinates of all constituents $\in B$.}where $B$ is the region of space complementary to $A$. 
The wave function\footnote{In the rest frame of the proton, it can be interpreted as the wave function of the incident photon.} corresponding to (\ref{vecspace}) can thus be written down in coordinate space as $\Psi_{AB}(y, z)$, and the corresponding density matrix $\rho_{AB} = |\Psi_{AB}\rangle \langle \Psi_{AB}|$ in coordinate space is 
\be\label{rho_pure}
\rho_{AB}(y, y', z, z') = \Psi_{AB}(y, z) \Psi_{AB}(y', z')^* .
\ee
The state described by (\ref{rho_pure}) is a pure quantum state with zero entropy.

Let us now introduce the density matrix describing the state probed in DIS. Since the region outside of $A$ is inaccessible to the measurement, we have to integrate over the coordinates $z \in B$ \cite{Landau}:
\be\label{rho_mixed}
\rho_A (y, y') = \int \Psi_{AB}(y, z) \Psi_{AB}(y', z)^* dz ,
\ee  
or in operator form, $\rho_A = \rm{tr}_B\ \rho_{AB}$.
The density matrix (\ref{rho_mixed}) describes a mixed state with a non-zero von Neumann entropy.
\vskip0.3cm

The Schmidt decomposition theorem \cite{Schmidt} (see \cite{Peres} for a discussion) states that the pure wave function  
$|\Psi_{AB} \rangle$ of our bi-partite system can be expanded as a {\it single} sum 
\be\label{schmidt}
|\Psi_{AB} \rangle = \sum_n \alpha_n |\Psi_{n}^A \rangle |\Psi_{n}^B \rangle 
\ee
for a suitably chosen orthonormal sets of states $|\Psi_n^A \rangle$ and $|\Psi_n^B \rangle$ localized in the domains $A$ and $B$, respectively, where $\alpha_n$ are positive and real numbers that are the square roots of the eigenvalues of matrix $C C^\dagger$.  In parton model, we assume that this full orthonormal set of states is 
given by the Fock states with different numbers $n$ of partons.

 The density matrix (\ref{rho_mixed}) of the mixed state probed in region $A$ can now be written down as  
\be
\rho_A = \rm{tr}_B\ \rho_{AB} = \sum_n \alpha_n^2\ |\Psi_n^A \rangle \langle \Psi_n^A |, 
\ee
where $\alpha_n^2 \equiv p_n$ is the probability of a state with $n$ partons. The identification of the basis $|\Psi_n^A \rangle$ in the Schmidt decomposition (\ref{schmidt}) with the states with a fixed number $n$ of partons is natural -- only in this case we do not have to deal with quantum interference between states with different numbers of partons, and such interference is absent in the parton model. Because the parton model represents a description of QCD that is a relativistic field theory, the number of terms in the sum (\ref{schmidt}) (the Schmidt rank) is in general infinite. Note that a pure product state with no entanglement would have a Schmidt rank one.
\vskip0.3cm
The von Neumann entropy of this state is given by 
\be\label{EE1}
S = - \sum_n\ p_n\ \ln p_n .
\ee
From our derivation it is clear that this entropy results from the entanglement between the regions $A$ and $B$, and 
can thus be interpreted as the entanglement entropy. In terms of information theory, Eq. (\ref{EE1}) represents the Shannon entropy for the probability distribution $(p_1, ..., p_N)$. 

%%%%%%%%%%%%%%%%%%%%%%%%%%%%%%%%%%%%%%%%%%%%%%%%%%%%
\begin{figure}
\begin{center}
\vspace{-1cm}
\includegraphics[width=6cm]{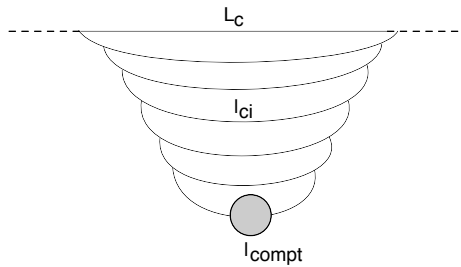}
\end{center}
\vspace{-2cm}
\caption{The parton cascade in deep inelastic electron-proton scattering. In the target rest frame, the partonic fluctuation develops over the longitudinal distance $L=(mx)^{-1}$, where $m$ is the proton mass. It interacts with the target that probes the partonic fluctuation with a resolution scale given by the proton's Compton wavelength $\epsilon = m^{-1}$.}
\label{casc}
\end{figure}

\vskip0.3cm

We will now evaluate the probabilities  $p_n$ and the corresponding entropy in two cases: i) a toy $(1+1)$ dimensional model of non-linear QCD evolution; and ii) in full $(3+1)$ dimensional case where the non-linear evolution is described by the Balitsky-Kovchegov (BK) equation \cite{KOLEB}. 
 
\subsection{1 + 1 toy model of non-linear QCD evolution}
%%%%%%%%%%%%%%%%%%%%%%%%%%%%%%%%%%%%%%%%%%%%%%%%%%%%%%%%
It will be convenient for us to describe the parton evolution using the dipole representation -- in this representation, a set of partons is represented by a set of color dipoles.  
In this section we consider a  $(1+1)$ dimensional toy model that emerges from the BK equation 
if one fixes the sizes of the interacting dipoles \cite{MUDI,LELU}. In this model the BFKL equation for the dipole scattering cross section $\sigma$ at a rapidity $Y$ is reduced to
\be \label{TM1}
\frac{d \sigma\Lb Y \Rb}{d Y} \,\,=\,\,\Delta  \, \sigma\Lb Y \Rb ,
\ee
where $\Delta$ is the BFKL intercept. The 
\eq{TM1} reproduces the power-like increase of the cross section with energy, $\exp(\Delta Y) = (1/x)^\Delta$.  

Let us now introduce  $P_n\Lb Y \Rb$, which is the probability to find $n$ dipoles (of a fixed size in our model) at  rapidity $Y$. For this probability we can write the following 
recurrent equation (see \fig{eq}):

%%%%%%%%%%%%%%%%%%%%%%%%%%%%%%%%%%%%%%%%%%%%%%%%%%%%
\begin{figure}
\begin{center}
\vspace{-5cm}
\includegraphics[width=12cm]{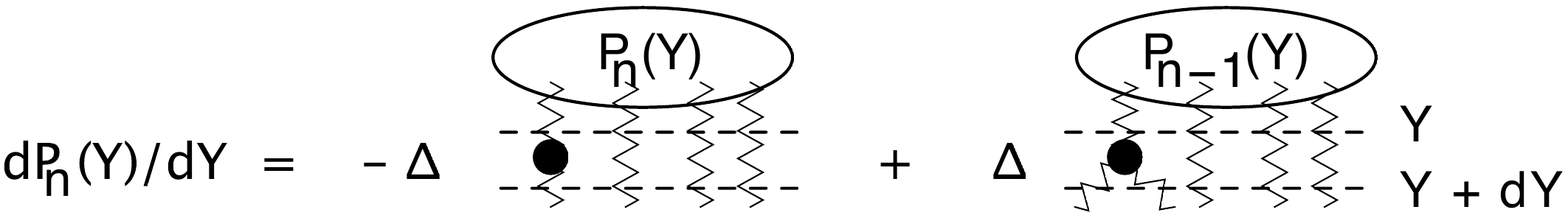}
\end{center}
\vspace{-7cm}
\caption{The graphic form of the equation (\ref{EQ}) for the probability of $n$ dipoles occurring at rapidity $Y$.}
\label{eq}
\end{figure}

%%%%%%%%%%%%%%%%%%%%%%%%%%%%%%%%%%%%%%%%%%%%%%%%%%%%
\be \label{EQ}
 \frac{d P_n\Lb Y\Rb}{d Y}\,\,=\,\,- \Delta\,n\,  P_n\Lb Y\Rb \,\,+\,\,\Lb n - 1 \Rb \Delta P_{n-1}\Lb Y .\Rb
\ee
This is a typical cascade equation in which the first term describes the depletion  of  the probability to find $n$ dipoles due to the splitting into $(n+1)$ dipoles, while the second one -- the growth due to the splitting of $(n-1)$ dipoles into $n$ dipoles. 

The \eq{EQ} can be re-written in a more convenient form by introducing the generating function
\be \label{TM2}
Z\Lb Y, u\Rb\,\,=\,\,\sum_n \,P_n\Lb Y\Rb \,u^n .
\ee
At the initial rapidity $Y=0$ we have only one dipole  so $P_1\Lb Y = 0 \Rb =1$ and $P_{n > 1} \,=\,0$ (so the state is pure);  at $u = 1$,  $Z\Lb  Y, u=1\Rb\,\,=\,\,\sum_n P\Lb y \Rb\,=\,1$. These two properties  determine the initial and the boundary conditions for the generating function:
\be \label{TM3}
Z\Lb Y = 0,u\Rb \,\,=\,\,u;~~~~~~~~~Z\Lb Y, u=1\Rb\,\,=\,\,1.
\ee
We can now re-write the \eq{EQ} as the following equation for the generating function:
\be \label{TM4}
\frac{\partial  Z\Lb Y, u\Rb}{ \partial Y}\,\,=\,\,- \Delta \, u \Lb 1 - u\Rb \frac{\partial Z\Lb Y, u\Rb}{\partial u} .
\ee
%%%%%
  It is instructive to observe that (\ref{TM4}) implies a non-linear equation for $Z\Lb Y,u\Rb$ \cite{MUDI,LELU}. Indeed, the general solution to (\ref{TM4}) is of the form $Z\Lb Y,u\Rb = Z(u(Y))$; if we substitute this function into (\ref{TM4}),  the derivatives $\partial Z/\partial u$ on the l.h.s. and r.h.s. of  (\ref{TM4}) cancel, and we get a differential equation for the function $u(Y)$. By using the first of the initial conditions (\ref{TM3}), we can then re-write   (\ref{TM4}) for rapidity near the one at which the initial condition is provided as 
\be \label{TM13}
  \,\,\frac{\partial\,Z}{\partial \,Y}\,\,=\,\,-\,\Delta 
\left(Z  \,\,-\,\, Z^2\right) .
\ee
Therefore, our parton cascade includes the interactions between the partons that lead to non-linear evolution in QCD\footnote{The dipole scattering amplitude in our model is  given by 
$N\Lb Y\Rb = 1 -  Z\Lb Y, 1 - \gamma\Rb$, where $\gamma$ is the dipole scattering amplitude  at $Y=0$. It also obeys the following non-linear equation : $d N\Lb Y\Rb/d Y\,=\,\Delta\Lb N - N^2\Rb$\cite{LELU,K}.}.

\begin{figure}[h]
\begin{center}
\includegraphics[width=9cm]{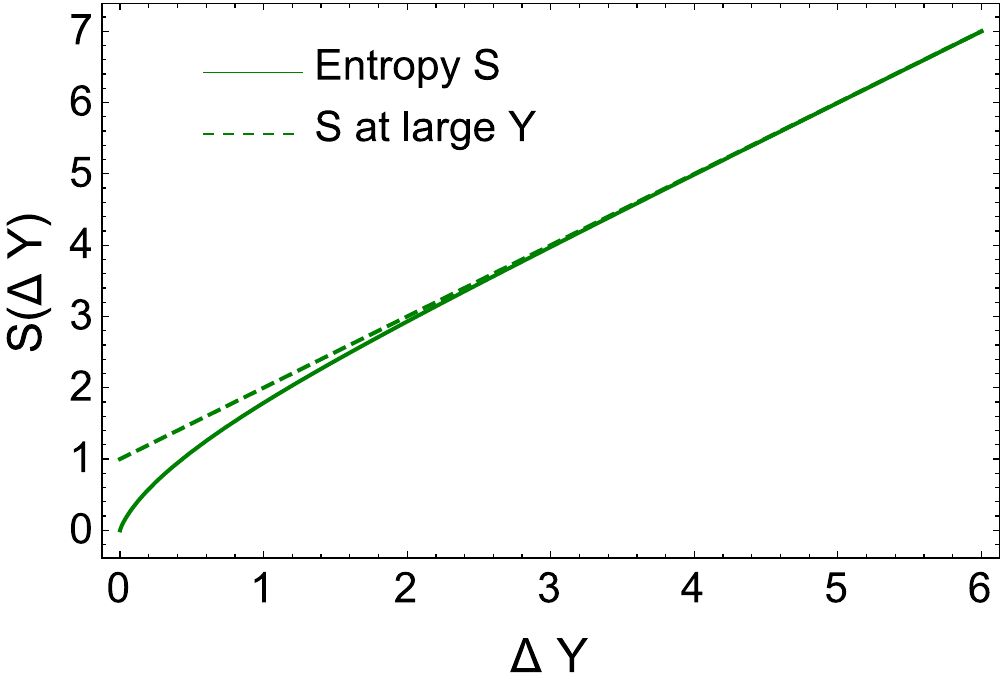}
\end{center}
\caption{The entanglement entropy a function $\Delta Y$ ($\Delta$ is the BFKL intercept). The dashed line corresponds to the large rapidity (or, equivalently, small $x$) limit of \eq{ent_smallx}.}
\label{S}
\end{figure}

The solution to (\ref{TM4}) with the initial and boundary condition of \eq{TM3} takes the form  \cite{LELU}
\be \label{SOL}
Z\Lb Y,\,u\Rb\,\,=\,\,\frac{u\,\,e^{\,-\,\Delta\, Y}}{1\,\,+\,\,u\,\,
\Lb e^{\,-\,\Delta\,Y}\,-\,1\Rb}\,\,=\,\,u\,e^{\,-\,\Delta\, Y}\,\sum^\infty_{n=1}\,u^n \Lb 1\,-\, e^{\,-\,\Delta\,Y}\Rb^n .
\ee
Comparing \eq{SOL} with \eq{TM2} one can see that
\be \label{TM5}
P_n\Lb Y \Rb\,\,=\,\,e^{\,-\,\Delta\, Y }\Lb 1\,-\, e^{\,-\,\Delta\,Y}\Rb^{n - 1} .
\ee
\vskip0.3cm
We are now in a position to calculate the von Neumann entropy of the system given by the Gibbs formula (\ref{EE1}) by identifying the probabilities of micro-states $p_n$ with the probabilities to find $n$ dipoles inside the hadron $P_n$ given by (\ref{TM5}), $p_n = P_n(Y)$.
The resulting entropy is given by
 \be \label{TM7}
 S\,\,=\,\,- \sum_n e^{\,-\,\Delta\, Y }\Lb 1\,-\, e^{\,-\,\Delta\,Y}\Rb^{n - 1}\Big( - \ln \Lb e^{\Delta Y }-1\Rb + n\, \ln\Lb 1 - e^{\,-\,\Delta\,Y}\Rb\Big) .
 \ee 
By using the generating function, \eq{TM7} can be re-written in the following way: 
 \be \label{TM8}
 S\Lb Y \Rb\,\,=\,\,\ln \Lb e^{\Delta Y }-1\Rb\,Z\Lb Y ,\,u = 1\Rb  \,\,+\,\,\ln\Lb \frac{1}{ 1 - e^{\,-\,\Delta\,Y}}\Rb \,u \frac{\partial Z\Lb Y ,\,u\Rb}{\partial u}\Bigg{|}_{u = 1} ,
 \ee
 which leads to 
 \be \label{TM9}
  S \Lb Y \Rb\,\,=\,\,\ln \Lb e^{\Delta Y }-1\Rb\,\,+\,\,e^{\Delta Y } \, \ln\Lb \frac{1}{ 1 - e^{\,-\,\Delta\,Y}}\Rb   .
  \ee 
  One can see that  at large $\Delta Y \,\gg\,1$\, 
  \be\label{ent_smallx}
  S (Y)  \,\to\, \Delta Y .
  \ee
  %%%%%%%%%%%%%%%%%%%%%%%%%%%%%%%%%%%%%%%%%%%%%%%%%%%%
In \fig{S} we show the dependence of the entropy on rapidity, as given by (\ref{TM9}) --  one can see that the asymptotic behaviour of \eq{ent_smallx} starts rather early, at $\Delta Y \simeq 2$.
    %%%%%%%%%%%%%%%%%%%%%%%%%%%%%%%%%%%%%%%%%%%%%%%%%%%%

\vskip0.3cm

To establish the relation between the entropy and the parton distribution, let us now evaluate the latter within the same framework.  We will define the parton distribution $xG(x)$ as the average number $ \langle n\rangle$ of partons at a given Bjorken $x$. 
 Using \eq{TM2} and \eq{SOL} we can calculate this number:
  \be \label{STRFUN1}
xG\Lb x\Rb\,=\,  \langle n\rangle \,\,=\,\,\sum_n n P_n\Lb Y\Rb\,=\,u \frac{d Z\Lb Y,u\Rb}{d u}\Bigg{|}_{u=1}\,=\,e^{\Delta\,Y} = \Lb \frac{1}{x}\Rb^\Delta .
  \ee
  
Note that in perturbative QCD  $xG\Lb x, Q\Rb$  obeys the BFKL evolution equation and grows at small $x$ as $(1/x)^\Delta$ with $\Delta = 4\,\ln 2\ \bas$, where $\bas\,\equiv\,\as\,N_c/\pi$. It should be stressed that the multiplicity of gluons ($xG\Lb x \Rb$)  evolves in accord with the linear evolution equation in spite of  the non-linear equation \eq{TM4}  for the generating function $Z$.
  \vskip0.3cm
  
  By comparing (\ref{STRFUN1}) and (\ref{ent_smallx}) we can see that at small $x$ the relation between the entropy and the structure function becomes very simple:
  \be \label{TM10}
  S\,\,=\,\,\ln\Lb xG\Lb x \Rb\Rb .
  \ee
  On the other hand for $\Delta Y < 1$ corresponding to a small increase in the number of partons, the entropy $S$ is given by
  \be \label{TM11}
    S\,\approx\,-\ln\left[\frac{xG\Lb x\Rb - xG\Lb x=x_0\Rb}{xG\Lb x=x_0\Rb}\right] \, \left[\frac{xG\Lb x\Rb - xG\Lb x=x_0\Rb}{xG\Lb x=x_0\Rb}  \right],
  \ee
  where $xG\Lb x=x_0\Rb$ is the number of partons at an initial value of $ x=x_0  $. In \eq{TM11} we assumed that $\delta xG  \equiv xG\Lb x\Rb - xG\Lb x=x_0\Rb \ll xG\Lb x=x_0\Rb$.
 \vskip0.3cm
 
It is important to note that the small $x$, large rapidity relation (\ref{TM10}) emerges in the limit where all probabilities $P_n$ become equal, see (\ref{TM5}) -- in this regime, an exponentially large number of partonic micro-states occur with equal and exponentially small probabilities $P_n(Y) = \exp(-\Delta Y) = 1/\langle n \rangle$.  It is well known that this equipartitioning of micro-states maximizes the von Neumann entropy and describes the maximally entangled state. We thus conclude that at small $x$ the proton represents a maximally entangled quantum state of partons.
\vskip0.3cm

In terms of information theory, the maximal value of the Shannon entropy (\ref{EE1}) achieved at small $x$ means that all ``signals" with different numbers of partons are equally likely, and it is impossible to predict how many partons will be detected. In other words, the information about the structure of the proton  (provided through an initial condition at some large $x_0$) becomes completely scrambled at small $x$ -- so non-linear QCD evolution is a very noisy ``communication channel" between large and small $x$. A particular realization of this ``information scrambling" is a chaotic behavior at small $x$ observed earlier \cite{Kharzeev:2005gn} in a discrete model of non-linear QCD evolution; however as we have shown the entropy attains it maximal value even if the chaos is not present. This implies that at sufficiently small $x$ the structure functions of all hadrons should become universal, i.e. independent of the initial conditions.
 
 \subsection{Multiplicity distribution}
 
 The information about the entropy of the final state is contained in the hadron multiplicity distribution. It is thus of interest to evaluate it assuming that it is the same as the parton multiplicity distribution that we have computed above. If the two distributions appear similar, it would suggest the absence of a substantial entropy increase during the transformation of partons to hadrons. Another reason for evaluating the multiplicity distribution stems from the fact that the entanglement entropy reaches its value (\ref{ent_smallx}) corresponding to a maximally entangled, equipartitioned state at a relatively modest rapidity difference $Y \leq 2/\Delta \simeq 6$ readily accessible at current hadron colliders.
It is important thus to check whether this approximately equipartitioned form of the entanglement entropy is consistent with the experimental hadron multiplicity distributions.  
\vskip0.3cm

Since the average multiplicity in our case is $ < n > \equiv \bar{n} \, =\, e^{\Delta Y}$ we can re-write the multiplicity distribution (\ref{TM5}) in the following form:
\be \label{MD}
P_n\Lb Y \Rb\,\,=\,\,e^{\,-\,\Delta\, Y }\Lb 1\,-\, e^{\,-\,\Delta\,Y}\Rb^{n - 1} \,\,=\,\,\frac{1}{\bar{n}}\Lb \frac{\bar{n} - 1}{\bar{n}}\Rb^{n - 1}\,=\,\frac{1}{\bar{N}}\Lb \frac{\bar{N}}{\bar{N} + 1}\Rb^n ,
\ee
where we have denoted $\bar{N} = \bar{n} - 1$.
Comparing \eq{MD} with the general form for the negative binomial distribution (NBD)
\be \label{NBD}
P^{\rm NBD}\Lb r, \bar{n},n\Rb\,\,=\,\,\Lb \frac{r}{r + 
\langle n \rangle}\Rb^r 
\frac{\Gamma\Lb n + r\Rb}{n!\, \Gamma\Lb r \Rb}\Lb 
\frac{\langle n \rangle}{r\,+\,\langle n \rangle}\Rb^n  ,
\ee
we see that our result \eq{MD} leads to the hadron multiplicity distribution that can be written down as
\be \label{MDNB}
P_n = \frac{\sigma_n}{\sigma_{\rm inel}}\,\,=\frac{\bar{n} -1}{\bar{n}}P^{\rm NBD}\Lb 1, \bar{n} - 1,n\Rb ,
\ee
where $\sigma_n$ is the cross section of producing $n$ hadrons in a collision, and $\sigma_{\rm inel}$ is the inelastic cross section. 
Therefore at large $\bar{n}$  our distribution is close to the negative binomial distribution with number of failures $r =1$ and with probability of success $p = \bar{N}/\Lb \bar{N} + 1\Rb = 1 - 1/\bar{n}$.
\vskip0.3cm

It turns out that the distribution given by \eq{MDNB} describes quite well the experimental distributions in high energy proton-proton collisions measured  at the LHC \cite{MDLHC}. For comparison with the experiment it is convenient to use the cumulants 
\be\label{cum_def}
C_q =\ <n^q>/<n>^q ,
\ee
 where $<\dots >$ denotes the average over the  distribution in hadron multiplicity $n$. These quantities can be readily computed using the generating function $Z$ given by (\ref{SOL}):
\be \label{CQ}
C_q\,\,=\,\,\Lb u \frac{d}{d u}\Rb^q Z\Lb Y, u\Rb\Bigg{|}_{u = 1} .
\ee
Using (\ref{SOL}) we can compute a few of the lowest cumulants up to $C_5$ that  which have been measured experimentally at the LHC at c.m.s. energy of $\sqrt{s} = 7$ TeV:
\bea\label{CQ1}
C_2 &=& 2 - 1/\bar{n};~~~~ C_3\,\,=\,\,\frac{ 6 (\bar{n} - 1)\bar{n} +1}{\bar{n}^2};\nn\\
C_4&=& \frac{(12 \bar{n}(\bar{n} - 1) +1)(2 \bar{n} -1)}{\bar{n}^3};~~~~~ C_5\,\,=\,\, \frac{(\bar{n} - 1)( 120 \bar{n}^2 (\bar{n} - 1) + 30 \bar{n} )+1}{\bar{n}^4} .
\eea

Using the experimental multiplicity in the rapidity window $|\eta | \leq 0.5$ equal to $\bar{n} = 5.8 \pm 0.1$ \cite{PDG} we get from (\ref{CQ1}) the following predictions for the cumulants: $C_2 \simeq 1.83$, $C_3 \simeq\,5.0$, $C_4 \simeq 18.2$ and $C_5 \simeq 83$.  These values are in a reasonably good agreement with the experimental data of Ref.\cite{MDLHC} (see Fig.6-b of that paper): $C_2^{\rm exp} = 2.0 \pm 0.05$, 
$C_3^{\rm exp} = 5.9 \pm 0.6$, 
$C_4^{\rm exp} = 21 \pm 2$, and 
$C_5^{\rm exp} = 90 \pm 19$. This agreement indicates that the multiplicity distribution of the produced hadrons is very close to the distribution in the number of partons that determines the entanglement entropy. 
\vskip0.3cm

It is instructive  to put  the upper bounds for these cumulants achieved at asymptotically high collision energy $\sqrt{s} \to \infty$, when the average multiplicity $\bar{n}$ becomes very large. Taking the limits of (\ref{CQ1}) at $\bar{n} \to \infty$ we get $C_2 = 2$, $C_3= 6$, $C_4=24$ and $C_5 = 120$ as a prediction for the asymptotically high energies. Comparing these numbers to the experimental values \cite{MDLHC} listed above, we see that the multiplicity distribution measured at $\sqrt{s} = 7$ TeV is already quite close to the expected asymptotic form.

  \subsection{Relation to the entanglement entropy in conformal field theory}\label{ent_sec}

At small $x$, the formulae (\ref{TM10}) and (\ref{STRFUN1}) yield the following result for the von Neumann entropy: 
\be\label{sb}
S(x) = \Delta \ln[1/x] = \Delta \ln \frac{L}{\epsilon},
\ee
where $L = (mx)^{-1}$ is the longitudinal distance probed in DIS ($m$ is the proton mass) and $\epsilon \equiv 1/m$ is the proton's Compton wavelength, see Fig.\ref{casc}. This expression looks very similar to the well known result for the entanglement entropy in  $(1+1)$ conformal field theory (CFT) \cite{HLW, CARDY}:
\be\label{se}
S_E = \frac{c}{3} \ \ln\frac{L}{\epsilon},
\ee
where $L$ is the length of the probed region, $\epsilon$ is the regularization scale describing the resolution of the measurement, and $c$ is the central charge of CFT that counts the number of degrees of freedom. The divergence in (\ref{se}) reflects the growth of the number of states near the boundary of the probed region when the resolution of the probe $\epsilon$ increases. 

The divergence in (\ref{sb1}) is precisely of the same origin -- the coherent quantum state of partons in DIS extends in the target rest frame over the distance $L = (mx)^{-1}$, and is probed with the resolution given by the proton's Compton wavelength $\epsilon \equiv 1/m$. The limit $\epsilon \to 0$ in (\ref{se}) is obviously equivalent to the small $x$ limit $L = (mx)^{-1} \to \infty$ in (\ref{sb1}), as in both cases $L/\epsilon \to \infty$.
\vskip0.3cm

Several comments on the possible relation between the results  (\ref{sb1}) and (\ref{se}) are in order:

First, (\ref{sb1}) refers to the quantum state, since the non-linear evolution equations that we used to derive (\ref{sb1}) is a faithful representation of RG flow in quantum field theory. The divergence of (\ref{sb1}) at small $x$ thus reflects the presence of the infinite number of states present in the theory. For the entire coherent quantum state, the entropy (\ref{sb1}) should vanish -- and it does: when the resolution of the measurement $\epsilon \equiv 1/m$ coincides with $L = (mx)^{-1}$ at $x=1$, $S \to 0$. This property is obviously shared between (\ref{sb1}) and (\ref{se}).

Second, in the limit when partons become incoherent the entropy (\ref{sb1}) should become extensive in $L$ -- for example, at high temperature $T \gg L^{-1}$ the entropy of one-dimensional gas is $S \sim L T$. The entanglement entropy (\ref{se}) in $(1+1)$ CFT has been evaluated also at finite temperature \cite{CARDY}, with the following result:
\be\label{finT}
S_E \sim \frac{c}{3} \ \ln\left(\frac{1}{\pi T \epsilon}\right)\ \sinh(\pi T L) .
\ee
At low temperatures $T \ll L^{-1}$ (\ref{finT}) reduces to (\ref{se}), whereas in the high temperature limit $T \gg L^{-1}$ we indeed obtain $S_E \sim c/3\ L T$ as expected for the extensive Boltzmann entropy of a one-dimensional gas.

\vskip0.3cm

  The similarity of (\ref{sb1}) and (\ref{se}) makes it plausible that at small $x$ the field theory describing parton evolution approaches a fixed point corresponding to a CFT with the central charge $c = 3 \Delta$. 
Let us discuss this in more detail. The length of the region $L = (mx)^{-1}$ probed in DIS grows at small $x$ (high energies), whereas $\epsilon \equiv 1/m$ stays fixed. To compare with the behavior in 2D field theory, we can however keep $L$ fixed and decrease the value of $\epsilon$ -- the high-energy behavior of our model is thus mapped onto the ultraviolet behavior of the 2D field theory. In other words, as the energy increases, we resolve shorter distances. The number of degrees of freedom increases from infrared to ultraviolet, and in $(1+1)$ field theory this intuitive expectation is confirmed by the rigorous $c-$theorem \cite{ZAMO} stating that the behavior of $c$ is monotonic under renormalization group flow.  

This allows us to conjecture a bound on the small $x$ behavior of the parton distributions. Indeed, in two dimensional CFTs the central charge assumes discrete values given by (see e.g. \cite{CARDY}):
\be
c = 1 - \frac{6}{m(m+1)}, \ \ m = 3,4,...,\infty .
\ee
The largest value of $c=1$ corresponds to the free bosonic field theory -- this is a likely fixed point in our theory as it corresponds to the asymptotically free behavior at short distances. Assuming that at small $x$ the partonic system indeed is described by a CFT, i.e. looks the same at all scales, we can thus put an upper bound on the value of $\Delta = c/3$: 
\be
\Delta \leq \frac{1}{3} .
\ee
The growth of parton multiplicities at high energies should thus be limited by
\be\label{bound_en}
xG(x) \leq {\rm const}\ \frac{1}{x^{1/3}} .
\ee
In fact, the value $\Delta = 1/3$ describes quite well the small $x$ behavior observed in DIS experiments \cite{PHEN1,PHEN2,PHEN3}\footnote{The experimental data on the deep inelastic structure function $F_2$ show that $F_2\Lb x, Q\Rb \propto \,\Lb \frac{1}{x}\Rb^\Delta$, with $\Delta $ that increases as a function of $Q^2$ from 0.2  to 0.35 for $Q^2 \, \geq\,5 \,{\rm GeV}^2$ (see Ref.\cite{PHEN2}). The modern fits of experimental data are based on  three  ingredients: non-linear evolution, running QCD coupling and next-to-leading order corrections to the BFKL kernel. Therefore, in general it is rather difficult to extract the value of effective value of $\Delta$ for the linear BFKL evolution. However, for large $Q^2 $ both non-linear and NLO corrections are rather small and the effective running QCD coupling turns out to be on the order of 0.1 (see Ref.\cite{PHEN1})  leading to $\Delta \approx 0.3$ \cite{PHEN1,PHEN2,PHEN3}.}.
 If we interpret this value in terms of the leading order BFKL result $\Delta = 4\,\ln 2\ \bas$, it implies the strong coupling value $\bas \simeq 0.12$. Of course, the relation of our result for the entanglement entropy (\ref{sb}) to the CFT one (\ref{se}) at this point is only a conjecture that will have to be verified. 
\vskip0.3cm

The ``asymptotic" small $x$ regime in which the formulae (\ref{TM10}), (\ref{sb1}) apply begins at $Y \leq 2/\Delta = 6$ (see \fig{S}), or at $x = \exp(-Y) \leq 10^{-3}$. It is accessible to the current and planned experiments, and can be investigated at the future Electron-Ion Collider (EIC). 
\vskip0.3cm

The small $x$ regime described by (\ref{TM10}) and (\ref{sb1}) implies the equipartitioning between the partonic micro-states that maximizes the entropy. It can thus be viewed as an analog of thermal equilibrium for the parton system at small $x$ -- just like statistical systems approach the thermally equilibrated macro-state with the largest entropy, small $x$ evolution leads to the universal state in which the entropy assumes the maximal value for a given $x$. It is thus natural to associate this regime with parton saturation corresponding to the equilibrium between the parton splitting and recombination processes.

\subsection{Entanglement entropy from the $(3+1)$ dimensional Balitsky-Kovchegov equation}
%%%%%%%%%%%%%%%%%%%%%%%%%%%%%%%%%%%%%%%%%%%%%%%%%%%%%%
  In this section we consider the evolution in $(3+1)$ dimensional QCD. We will see that the result for the entropy in this case is very similar to the one obtained above in the $(1+1)$ toy model. As discussed in Refs.\cite{MUDI,LELU}  the parton cascade equation \eq{EQ} in the $(3+1)$ case can be written down in the following form:
  \bea  \label{PC1}
\frac{\partial\,P_n\left(Y;\,r_1,\,r_2 \dots r_i 
\dots r_n \right)}{ 
\bar{\alpha}_s\,\partial\,\Lb Y \Rb}\,&=&\,-\,
\sum^n_{i=1}\,\omega(r_i) \,
P_n\left(Y ;\,r_1,\,r_2 \dots r_i \dots r_n 
\right)\nonumber \\
&+&\,\,\sum^{n-1}_{i=1} \,\frac{(\vec{r}_i\,+\, 
\vec{r}_n)^2}{(2\,\pi)\,r^2_i\,r^2_n}\,
P_{n - 1}\left(Y ;\,r_1,\,r_2
\dots  (\vec{r}_i \,+\, \vec{r}_n)\dots r_{n-1} \right)
\eea
  where $P_n\Lb Y ; \{r_i\}\Rb$ is the probability to have $n$-dipoles with size $r_i$ at rapidity $Y-y$.
  This QCD  cascade leads to Balitsky-Kovchegov equation \cite{KOLEB} for the  amplitude and gives the theoretical description of the DIS.
  
  Comparing \eq{PC1} with \eq{EQ} we see that the probability for one dipole to survive is not a constant as in \eq{EQ} but depends on the dipole size: 
  \be \label{PC2}
 \bas \,\,\omega(r_i)\,\,\equiv\,\bas\ \omega_i\,\,=\,\,\frac{ 
\bas}{2\,\pi}\,\int_\rho 
\,\frac{r_i^2}{(\vec{r_i}\,-\,\vec{r}')^2\,r'^2}\,d^2 r'\,\,=\,\,
\bas \,\,\ln(r_i^2/\rho^2) ,
\ee 
where  $\rho$ is an infrared cutoff and $\bas\,=\,\as\,N_c/\pi$.  
The probability for a dipole of the size $|\vec{r}_1\,+\,\vec{r}_2|$ to decay into 
two  with the sizes $r_1$ and $r_2$ is equal to \cite{MUDI}
\be 
K\Lb r_1,r_2| \vec{r_1}+\vec{r}_2\Rb\,\,=\,\,\frac{\bas}{2\,\pi} \,\,\frac{(\vec{r_1} \,\,+\,\, 
\vec{r_2})^2}{r_1^{2}\,r_2^2} .
\ee

For $n=1$ \eq{PC1} has the solution
\be \label{N1}
P_1\Lb Y ; r_1\Rb\,=\,\delta\Lb \vec{r}-\vec{r}_1\Rb\,e^{ - \omega\Lb r_1\Rb \bas \Lb Y \Rb}
\ee
which reflects the fact that at $Y-y=0$ we have only one dipole of size $r$.  It means also that $P_{n>1}\Lb Y  = 0; \{r_i\}\Rb \,=\,0$.

Since $P_n\Lb Y ; \{r_i\}\Rb$ is the probability to find dipoles $\{r_i\}$, we have the following sum rule 

\be \label{SUMRU}
\sum_{n=1}^\infty\,\int \prod^n_{i=1} d^2 r_i \,P_n\Lb Y ; \{r_i\}\Rb\,\,=\,\,1 ,
\ee
i.e. the sum of all probabilities is equal to 1.

Replacing $P_n\Lb Y ; \{r_i\}\Rb$ by its Mellin image $P_n\Lb \omega; \{r_i\}\Rb$
\be \label{PC4}
P_n\Lb Y  ; \{r_i\}\Rb\,=\,\,\int^{\epsilon + i \infty}_{\epsilon - i \infty}\frac{d \omega}{2 \,\pi}e^{\omega\,\bas \, Y }\,P_n\Lb \omega; \{r_i\}\Rb
\ee
we reduce \eq{PC1} to the form
\be \label{PC5}
P_n\left(\omega; \{r_i\} \right)\,=\,-\,
\sum^n_{i=1}\,\omega_i  \,
P_n\Lb\omega; \{r_i\}\Rb\,+\,\,\,\frac{1}{2 \,\pi}\sum^{n-1}_{j=1} \,\frac{(\vec{r}_j\,+\, 
\vec{r}_n)^2}{\,r^2_j\,r^2_n}\,
P_{n - 1}\left(\omega;\,\{r_i,\vec{r}_j \to 
(\vec{r}_j \,+\, \vec{r}_n)\} \right) .
\ee
One can see that $P_n\Lb \omega,\{r_i\}\Rb$ can be re-written as
\be \label{PC6}
P_n\left(\omega; \{r_i\} \right)\,\,=\,\,2 \,\pi \,r^2\, \delta\Lb \vec{r} - \vec{r}_1\Rb\,\Lb \frac{1}{ 2 \,\pi}\Rb^n\prod_{i=1}^n\frac{1}{r^2_i}\,\Omega_n\Lb \omega,\{ \omega_i\}\Rb ,
\ee
with the following equation  for $\Omega_n\Lb \omega,\{ \omega_i\}\Rb$:
\be \label{OMEGA}
\omega\,\Omega_n\Lb \omega,\{ \omega_i\}\Rb\,=\,-\,\Lb \sum_{i=1}^n\,\omega_i\Rb \,\Omega_n\Lb \omega,\{ \omega_i\}\Rb\,\,+\,\,\sum^{n-1}_{j=1} \Omega_{n-1}\Lb \omega,\{ \omega_i,\omega_{jn}\}\Rb .
\ee
The solution is given by the  recurrent equation 
\be \label{OMEGA1}
\Omega_n\Lb \omega,\{ \omega_i\}\Rb\,\,=\,\,\Lb n-1\Rb\Omega_{n-1}\Lb \omega,\{ \omega_i,\omega_{n-1,n}\}\Rb\frac{1}{\omega \,+\, \sum^{n}_{j=1} \omega_j} ;
\ee
in writing \eq{OMEGA1} we used the symmetry between $r_i$ and introduced the following short notations: $\omega_i =\omega\Lb \vec{r}_i\Rb$ and $\omega_{ij} = \omega\Lb \vec{r}_i \,+\,\vec{r}_{j}\Rb$. Therefore, in \eq{OMEGA} and \eq{OMEGA1} $\omega_{jn} =  \omega\Lb \vec{r}_j \,+\,\vec{r}_{n}\Rb$ and $\omega_{n-1,n}\,=\,\omega\Lb \vec{r}_{n-1} \,+\,\vec{r}_{n}\Rb$.

We cannot solve \eq{OMEGA1} in an explicit way due to the term $\omega_{n-1,n}$. However, we are able to do this for two instructive cases. The first case corresponds to the double log approximation of perturbative QCD in which $r_i$ and $r_n$ are both  
larger than $|\vec{r}_i + \vec{r}_n|$. In this case \eq{OMEGA1} can be re-written as
\be \label{OMEGA11}
\Omega_n\Lb \omega,\{ \omega_i\}\Rb\,\,=\,\,\Lb n-1\Rb\Omega_{n-1}\Lb \omega,
\{ \omega_i\}\Rb\frac{1}{\omega \,+\, \sum^{n-2}_{j=1} \omega_j + 2 \omega_{n-1}} .
\ee
The general solution to \eq{OMEGA11} takes the form
\be \label{OMEGA12}
\Omega_n\Lb \omega,\{ \omega_i\}\Rb\,\,=\,\,\Lb n-1\Rb! \prod^{n}_{j=2} \frac{1}{\omega \,+\, \sum^{j-2}_{l=1} \omega_l + 2 \omega_{n-1}} .
\ee

The second case describes the decay of the large dipole into an asymmetric pair of dipoles, one large and one small.
In this case,  $|\vec{r}_i + \vec{r}_n|\,\to\,r_i$ while $r_n \,\ll\,r_i$.  As noted in Ref.\cite{LETU} this case corresponds to summation of terms  $\ln^n\Lb r^2_i Q^2_s\Rb$ for $ r^2_i Q^2_s \,\gg\,1$; in other words, it describes the behavior of the parton cascade deep inside of the saturation region. In this kinematic region 
$\omega_i = \ln\Lb r^2_i Q^2_s\Rb\,\equiv\,z_i$.  
The solution of \eq{OMEGA1} in this case takes the following form:

\be \label{OMEGA13}
\Omega_n\Lb \omega,\{ \omega_i\}\Rb\,\,=\,\,\Lb n-1\Rb! \prod^{n}_{j=1}\frac{1}{\omega \,+\, \sum^{j}_{l=1} \omega_l }\,\,=\,\,\Lb n-1\Rb! \prod^{n}_{j=1}\frac{1}{\omega \,+\, \sum^{j}_{l=1} z_l } .
\ee
Using \eq{PC4} and \eq{PC6} we find
\be \label{OMEGA14}
P_n\Lb Y  ; \{r_i\}\Rb\,=\,\,2 \,\pi \,r^2\, \delta\Lb \vec{r} - \vec{r}_1\Rb\,\Lb \frac{1}{ 2 \,\pi}\Rb^n\prod_{i=1}^n\frac{1}{r^2_i}\, \int^{\epsilon + i \infty}_{\epsilon - i \infty}\frac{d \omega}{2 \,\pi}e^{\omega\,\bas \, Y }\,\Omega_n\Lb \omega,\{ \omega_i\}\Rb .
\ee
For $\int \prod^n_{i=1} d^2 r_i \,P_n\Lb Y ; \{r_i\}\Rb$ we can re-write \eq{OMEGA14}  in the form
\be \label{OMEGA15}
\int \prod^n_{i=1} d^2 r_i \,P_n\Lb Y ; \{r_i\}\Rb\,=\,\,\int^{\epsilon + i \infty}_{\epsilon - i \infty}\frac{d \omega}{2 \,\pi}e^{\omega\,\bas \, Y }\int \prod^n_{i=1} \,d z_i \,\Omega_n\Lb \omega,\{ z_i\}\Rb .
\ee
Using Feynman parameters we can  simplify \eq{OMEGA14} as follows:
\bea \label{OMEGA151}
&&\int^{\epsilon + i \infty}_{\epsilon - i \infty}\frac{d \omega}{2 \,\pi}e^{\omega\,\bas \, Y } \,\Omega_n\Lb \omega,\{ z_i\}\Rb
\,=\,\nn\\
&&=\,\Lb \bas Y\Rb^n \int^1_0 \prod^n_{i=2}\, d \alpha_i \exp\Bigg\{ - \Lb z_1 + z_2\sum^n_{i=2} \alpha_i + z_3 \sum^n_{i=3} \alpha_i + \,\dots \,+\,z_l \sum^n_{i=l} \alpha_i\,+\, \dots\,+\,z_n\alpha_n \Rb \bas\,Y\Bigg\}\nn\\
&&= \,\,\Lb \bas Y\Rb^n \int^1_0 \prod^n_{i=2}\, d \alpha_i \exp\Bigg\{ - \Lb z_1 + \alpha_n\sum^n_{i=2}z_i + \alpha_{n-1} \sum^n_{i=3} z_i + \,\dots \,+\,\alpha_l \sum^n_{i=l} z_i\,+\, \dots\,+\,z_n\alpha_n \Rb \bas\,Y\Bigg\}\nn\\
&&=\,\, e^{ - \bas \,z_1\,Y}\prod^n_{i=2}\Lb \frac{ 1 - e^{- \Lb \sum^n_{l=i}\,z_l\Rb\,\bas\,Y}}{\sum^n_{l=i}\,z_l}\Rb\,\,\equiv\,\,\Lb \bas \,Y\Rb^n e^{ - \bas \,z_1\,Y}\prod^n_{i=2}  \Phi\Lb \bas Y\, \sum^n_{l=i}\,z_l\Rb\eea
\eq{OMEGA15} now takes the form
\bea \label{OMEGA16}
&&\int \prod^n_{i=1} d z_i \,P_n\Lb Y ; \{z_i\}\Rb\,=\nn\\
&&=\,e^{- \bas\,z_1\,Y}\int^{\bas\,z_1\,Y}_0\,\Phi\Lb t_n\Rb \,d t_n\,\int^{t_n}_0 d t_{n-1}\,\Phi\Lb t_{n-1}\Rb\,\dots\,\int^{t_3}_0 d t_2\,\Phi\Lb t_2\Rb
\,=\,\,\frac{1}{n!}\Xi^n\Lb \bas\,z_1\,Y\Rb\,e^{- \bas\,z_1\,Y} ,
\eea
where $t_i = \bas Y\, \sum^n_{l=i}\,z_l$ and
\be \label{XI}
\Xi\Lb t \Rb\,=\,\int^t_0 \Phi\Lb t'\Rb d t'\,=\,C\,+\,\Gamma\Lb 0, t\Rb\,+\,\ln t ;
\ee
here $C$ is the Euler constant and $\Gamma\Lb 0,t\Rb$ is the incomplete gamma function. One can see that
\be \label{XI1}
 \Xi\Lb t \Rb \,\,= \,\,\left\{ \begin{array}{ll}
t &\,\, \mbox{if $t \ll 1$};\\
\ln\Lb 1/t\Rb & \,\,\mbox{if $t \gg 1$}.\end{array} \right. 
 \ee
In \eq{OMEGA16} we used the usual ordering condition: 
\be \label{ORDR}
z_1 \,\gg\,z_2\,\gg\,\dots\,\gg\,z_i\,\gg\,z_{i-1}\,\gg\,\dots\,\gg\, 0 .
\ee
From
\eq{OMEGA16} one can see that $\Omega$ satisfies the initial condition $\Omega_{n>1}\Lb Y - y = 0,\{r_i\}\Rb = 0$ since $\Xi^n\Lb \bas\,z_1\,Y\Rb\,\to\,0$ at $Y \to 0$.

%%%%%%%%%%%%%%%%%%%%%%%%%%%%%%%%%%%%%%%%%%%%%%%%%%%%
\begin{figure}[h]
\begin{center}
\includegraphics[width=9cm]{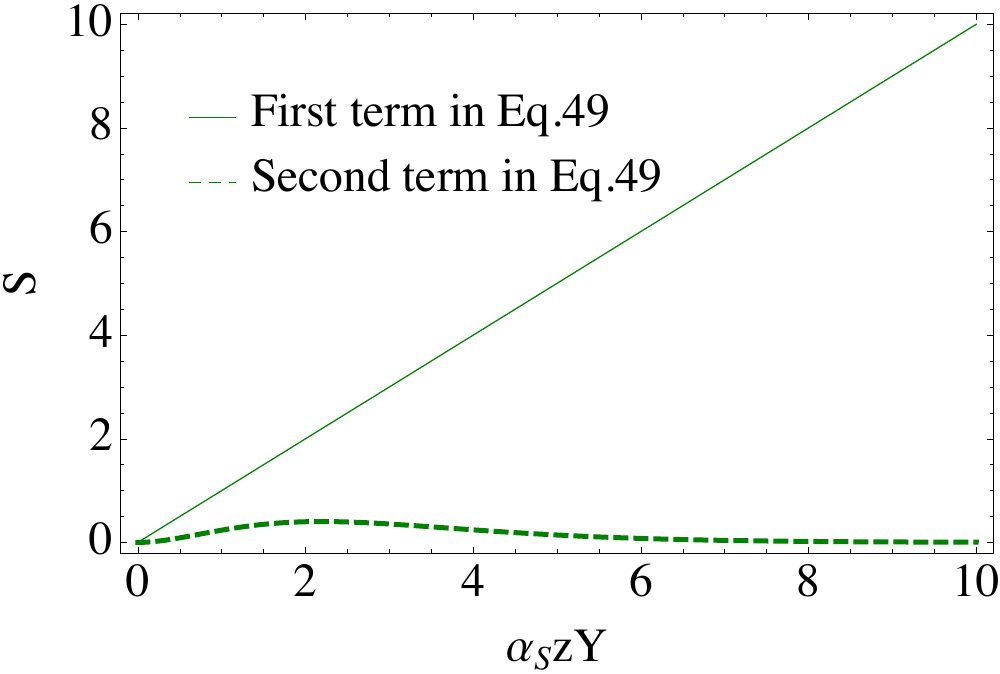}
\end{center}
\caption{Entropy dependence on $\bas \,z\,Y$ (see text). Solid line corresponds to the first tern in \eq{PCS4} while the dashed line describes the second term.}
\label{ss}
\end{figure}
%%%%%%%%%%%%%%%%%%%%%%%%%%%%%%%%%%%%%%%%%%%%%%%%%%%% 
To find the entropy we need to evaluate the Gibbs formula

\be \label{PCS1}
S\,\,=\,\,- \sum^\infty_{n=1}\prod^n_{i=1} \,\int d^2 r_i\,P_n\Lb Y ; \{r_i\}\Rb\,\ln\Big(P_n\Lb Y ; \{r_i\}\Rb\Big)
\ee
It is hard to calculate the integrals in \eq{PCS1} in general case, 
but fortunately we can find the entropy at large $Y $. Indeed, at  large values of $Y $ ($\omega_1\, \bas\,  Y \,\gg\,1$) $P_n $ reduces to the following form (see \eq{OMEGA151})
\be \label{PCS2}
P_n\Lb \omega,\{ \omega_i\}\Rb\,\,\xrightarrow{\omega_1\, \bas\,  Y \,\gg\,1}\,\,\,\,
e^{ - \omega_1 \bas  Y }\,e^{- \sum^n_{i=1} \,z_i}\Bigg\{\prod^n_{j=2}\frac{1}{\sum^{j}_{l=2}z_l}\,\,+\,\,
\dots\Bigg\} .
\ee
Plugging \eq{PCS2} into \eq{PCS1} we obtain
\bea \label{PCS3}
S\,&=&\,\omega(r) \bas Y \sum_{n=1}^\infty\,\int \prod^n_{i=1} d^2 r_i \,P_n\Lb Y - y; \{r_i\}\Rb\nn\\
\,&-&\,e^{ -  \bas z_1 Y }\,\sum^\infty_{n=1}\int \prod^n_{i=2} d z_i \,\Big\{ \sum^n_{i=2} z_i - \sum^n_{i=2} \ln\Lb \sum^n_{l=i} z_l\Rb\Big\}\,
\,\prod^n_{i=2}\Lb \frac{ 1 - e^{- \Lb \sum^n_{l=i}\,z_l\Rb\,\bas\,Y}}{\sum^n_{l=i}\,z_l}\Rb .
\eea
Using \eq{SUMRU} and neglecting $\sum^n_{i=2} \ln\Lb \sum^n_{l=i} z_l\Rb$ in \eq{PCS3} we reduce this equation to the form
\bea \label{PCS4}
S\,&=&\, \bas\,z\, Y\,
\,-\,e^{ -z \bas \Lb Y \Rb}\,\sum^\infty_{n=1}\int^{\bas z\,Y}_0 t_n\,d t_n \,\Phi\Lb t_n\Rb \,\frac{1}{(n-1)!} \Xi^{n-1}\Lb t_n\Rb\,\nn\\
&=&\, \bas\,z\, Y\,
\,-\,e^{ -z \bas \Lb Y \Rb}\int^{\bas z\,Y}_0 t_n\,d t_n\,\Phi\Lb t_n\Rb\exp\Lb \Xi\Lb t_n\Rb\Rb
\eea
with $z \equiv \ln\Lb r^2 Q^2_s\Rb$. We emphasize that the normalization of the first term is fixed by the condition  (\ref{SUMRU}) and does not depend on the approximations we make in evaluating the integrals over $r_i (z_i)$. 

The second term in \eq{PCS4} is small in comparison with the first one (see \fig{ss}).
 Therefore, the entropy at large $Y$ takes the form
\be \label{PCS5}
S\,\approx\, \bas \,z\,Y\ee
which coincides with the expression (\ref{ent_smallx}) that we obtained for the toy model in the previous section if we take $\Delta = \bas\ z = \bas \ln\Lb r^2 Q^2_s\Rb$. The characteristic dipole size $r^2$ in deep inelastic scattering is set by the momentum transfer $Q^2$. 

\section{Discussion}

Using non-linear evolution equations of QCD, we have evaluated the von Neumann entropy of the system of partons that is resolved in a DIS measurement at a given Bjorken $x$ and momentum transfer $q^2 = - Q^2$ (note that in  our estimates presented in the previous section $r^2 \sim 1/Q^2$). We have found that at small $x$ the relation between the entropy and the parton distribution becomes very simple and is given by (\ref{TM10}). 
In this small $x$ regime (corresponding to $x < 10^{-3}$ and rapidity $Y > 6$, as we estimated above), all partonic micro-states have equal probabilities -- the proton is composed by an exponentially large number $\exp(\Delta Y)$ of partonic micro-states that occur with equal and exponentially small probabilities $\exp(-\Delta Y)$. In this equipartitioned state, the entropy (\ref{EE1}) (that we have interpreted as resulting from the entanglement) is maximal -- so {\it the partonic state at small $x$ is maximally entangled}.

\vskip0.3cm

If we interpret (\ref{EE1}) in terms of information theory as the Shannon entropy, then the equipartitioning in the Maximally Entangled State (MES) means that all ``signals" with different number of partons are equally likely, and it is impossible to predict how many partons will be detected in a given event. In other words, the information about the structure of the proton encoded in an initial condition becomes completely scrambled in the MES at small $x$. Therefore the structure functions at sufficiently small $x$ should become universal for all hadrons.

\vskip0.3cm
Since the parton distribution and the entanglement entropy at small $x$ are related by  (\ref{TM10}), one may question the utility of entropy in characterizing the process of DIS. However there are several reasons to believe that the entanglement entropy is a useful DIS observable:
\begin{itemize}
\item{Identifying the entropy of partonic system as the entanglement entropy explains  the apparent loss of quantum coherence in the parton model, solving an old conceptual problem described in the Introduction. The entropy that we have found originates from the entanglement between the spatial domain probed by DIS and the rest of the target, 
whereas the entire proton is in a pure quantum state with zero entropy.
}
\item{Parton distributions have a well-defined meaning only for weakly coupled partons at large momentum transfer $Q^2$ -- but the entanglement entropy is a universal concept that applies to states at any value of the coupling constant. 
}
\item{Unlike the parton distributions, the entanglement entropy is subject to strict bounds -- for example, if the small $x$ regime is described by a CFT, the growth of parton distributions should be bounded by $xG(x) \leq {\rm const}\ x^{-1/3}$, see (\ref{bound_en}).  
}
\item{If the second law of thermodynamics applies to entanglement entropy (and there are indications \cite{Hubeny:2007xt} that it does), then the entropy of a final hadronic state $S_h$ cannot be smaller than the entropy $S(x)$ accessed at a given Bjorken $x$, and we expect the proportionality $S_h \sim S(x)$. The correspondence between the number of partons in the initial state and the number of hadrons in the final state is in accord with the ``parton liberation" \cite{Mueller:1999fp} and ``local parton-hadron duality" \cite{Dokshitzer:1987nm} pictures. The link between the entropy in the initial state at small $x$ and the final state entropy  has also been discussed in refs \cite{Baier:2000sb,Kharzeev:2000ph,Kharzeev:2001gp,Kharzeev:2005iz,Kharzeev:2006zm,Fries:2008vp}. 
}
\end{itemize}
\vskip0.3cm
 
 The entropy is a useful measure of information that can be obtained in an experiment -- therefore it is an appropriate general characterization of the outcome of a DIS measurement.  The entropic approach proposed here underlines the importance of measuring the hadronic final state of DIS. We thus encourage experimentalists to combine the measurements of the DIS cross sections with the determination of hadronic final state. The determination of the Shannon entropy of hadrons in the final state of DIS can be done using the event-by-event multiplicity measurements, see e.g. \cite{Bialas:1999wi,Atayan:2005cv}. 
As we estimated above, the ``asymptotic" small $x$ regime in which the formulae (\ref{TM10}), (\ref{sb1}) begins at $x \leq 10^{-3}$. It is accessible to the current and planned experiments, and can be investigated at the future Electron-Ion Collider (EIC) \cite{Accardi:2012qut}. 
 \vskip0.7cm

    We thank our colleagues at BNL, Stony Brook University, Tel Aviv University and UTFSM for
 stimulating discussions. We are grateful to A.H. Mueller and Yuri Kovchegov for positive and useful conversations. This work was supported in part by the U.S. Department of Energy under Contracts No.
DE-FG-88ER40388 and DE-AC02-98CH10886, BSF grant   2012124,   
 Proyecto Basal FB 0821(Chile),  Fondecyt (Chile) grant  
 1140842, and by CONICYT grant PIA ACT1406.

    \end{document}